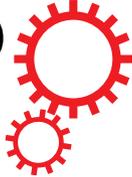

# Exact quantum Bayesian rule for qubit measurements in circuit QED

Wei Feng[1,2], Pengfei Liang[2], Lupei Qin[2] & Xin-Qi Li[2]

Developing efficient framework for quantum measurements is of essential importance to quantum science and technology. In this work, for the important superconducting circuit-QED setup, we present a rigorous and analytic solution for the effective quantum trajectory equation (QTE) after polaron transformation and converted to the form of Stratonovich calculus. We find that the solution is a generalization of the elegant quantum Bayesian approach developed in arXiv:1111.4016 by Korotokov and currently applied to circuit-QED measurements. The new result improves both the diagonal and off-diagonal elements of the qubit density matrix, via amending the distribution probabilities of the output currents and several important phase factors. Compared to numerical integration of the QTE, the resultant quantum Bayesian rule promises higher efficiency to update the measured state, and allows more efficient and analytical studies for some interesting problems such as quantum weak values, past quantum state, and quantum state smoothing. The method of this work opens also a new way to obtain quantum Bayesian formulas for other systems and in more complicated cases.

Continuous quantum weak measurement, which stretches the "process" of the Copenhagen's instantaneous projective measurement, offers special opportunities to control and steer quantum state[1,2]. This type of measurement is essentially a process of real-time monitoring of environment with stochastic measurement records and causing back-action onto the measured state. However, remarkably, the stochastic change of the measured state can be faithfully tracked. Because of this unique feature, quantum continuous weak measurement can be used, for instance, for quantum feedback of error correction and state stabilization, generating pre- and post-selected (PPS) quantum ensembles to improve quantum state preparation, smoothing and high-fidelity readout, and developing novel schemes of quantum metrology[1,2].

On the other aspect, the superconducting circuit quantum electrodynamics (cQED) system[3–5] is currently an important platform for quantum measurement and control studies[6–15]. In particular, the continuous weak measurements in this system, as schematically shown in Fig. 1(a), have been demonstrated in experiments[12–14], together with feedback control[15].

For continuous weak measurements, the most celebrated formulation is the quantum trajectory equation (QTE) theory, which has been broadly applied in quantum optics and quantum control studies[1,2]. However, in some cases of real experiment, numerical integration of the QTE is not efficient and alternatively, the *one-step* quantum Bayesian approach has been employed to update the quantum state based on the *integrated* output currents[12–15], as illustrated in Fig. 1(b). Meanwhile, one may notice that the Bayes' formula is the fundamental tool for classical noisy measurements[2]. In this work we carry out the analytic and exact solution of the effective QTE of the cQED system. Desirably, we find that the solution is a generalization of the elegant quantum Bayesian approach developed in ref. 16 by Korotkov and currently applied to circuit-QED experiments[13–15]. The new result is not bounded by the "bad"-cavity and weak-response limits as in ref. 16, and promises higher efficiency than numerically integrating the QTE. The Bayesian rule also allows more efficient and analytical studies for some interesting problems such as quantum weak values[17,18], past quantum state[19–21], and quantum state smoothing[22,23], etc.

## Results

**Exact Bayesian rule for cQED measurement.** We consider the cQED architecture consisting of a superconducting transmon qubit *dispersively* coupled to a waveguide cavity, see Fig. 1(a). The qubit-cavity interaction is given by the Hamiltonian[3–5], $H_{int} = \chi a^\dagger a \sigma_z$, where $\chi$ is the dispersive coupling rate, $a^\dagger$ and $a$ are respectively the creation and annihilation operators of the cavity mode, and $\sigma_z$ is the qubit Pauli operator. This interaction Hamiltonian characterizes a qubit-state-dependent frequency shift of the cavity which is used to perform quantum state measurement. After a qubit-state-dependent displacement transformation (called also "polaron" transformation) to eliminate the cavity degrees of freedom[24], the single quadrature homodyne current can be reexpressed as

$$I(t) = -\sqrt{\Gamma_{ci}(t)} \langle \sigma_z \rangle + \xi(t). \quad (1)$$

[1]Department of Physics, Tianjin University, Tianjin 300072, China. [2]Center for Advanced Quantum Studies and Department of Physics, Beijing Normal University, Beijing 100875, China. Correspondence and requests for materials should be addressed to X.-Q.L. (email: lixinqi@bnu.edu.cn)

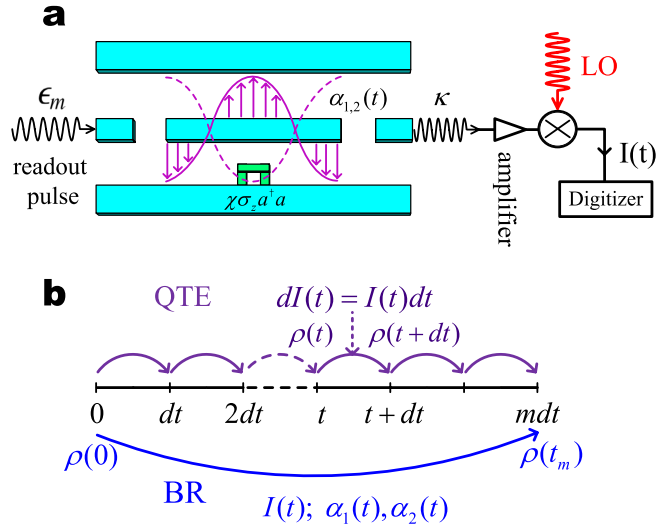

**Figure 1. Qubit measurement in circuit QED and the idea of one-step Bayesian state inference.** (**a**) Schematic plot for the circuit QED setup and measurement principle of qubit states via measuring the quadrature of the cavity field. The superconducting qubit couples dispersively to the cavity through Hamiltonian $\chi a^\dagger a \sigma_z$ which, under the interplay of cavity driving and damping, forms cavity fields $\alpha_1(t)$ and $\alpha_2(t)$ corresponding to qubit states $|1\rangle$ and $|2\rangle$. The leaked photon (with rate $\kappa$) is detected using the homodyne technique by mixing it with a local oscillator (LO). (**b**) Illustrative explanation for the advantage of the one-step Bayesian rule (BR) over the continuous (multi-step) integration of the quantum trajectory equation (QTE). That is, using the known functions $\alpha_{1,2}(t)$ and the detected current $I(t)$ of the homodyne measurement, the BR allows a one-step inference for the qubit state.

In this result, $\xi(t)$, originated from the fundamental quantum-jumps, is a Gaussian white noise and satisfies the ensemble-average properties of $E[\xi(t)] = 0$ and $E[\xi(t)\xi(t')] = \delta(t-t')$. $\Gamma_{ci}(t)$ is the *coherent* information gain rate which, together with the other two, say, the no-information back-action rate $\Gamma_{ba}(t)$ and the overall measurement decoherence rate $\Gamma_d(t)$, is given by[24]

$$\Gamma_{ci}(t) = \kappa |\beta(t)|^2 \cos^2(\varphi - \theta_\beta), \quad (2)$$

$$\Gamma_{ba}(t) = \kappa |\beta(t)|^2 \sin^2(\varphi - \theta_\beta), \quad (3)$$

$$\Gamma_d(t) = 2\chi \, \mathrm{Im}[\alpha_1(t)\alpha_2^*(t)]. \quad (4)$$

Here we have denoted the local oscillator's (LO) phase in the homodyne measurement by $\varphi$, the cavity photon leaky rate by $\kappa$, and $\beta(t) = \alpha_2(t) - \alpha_1(t) \equiv |\beta(t)| e^{i\theta_\beta}$ with $\alpha_1(t)$ and $\alpha_2(t)$ the cavity fields associated with the respective qubit states $|1\rangle$ and $|2\rangle$.

In this work, for brevity, we directly quote the transformed QTE results from ref. 24 where the derivation and explanation of the above rates can be found with details. Briefly speaking, the coherent information gain rate, $\Gamma_{ci}$, describes the backaction of information gain during the measurement, which corresponds to the "spooky" backaction rate termed in ref. 16 by Korotkov. $\Gamma_{ba}$, corresponding to the "realistic" backaction rate in ref. 16, characterizes the backaction of the measurement device not associated with information gain of the qubit state. And the overall decoherence rate, $\Gamma_d$, describes the ensemble average effect of the measurement on the qubit state, over large number of quantum trajectories.

Within the framework of "polaron" transformation, the effective quantum trajectory equation (QTE) for qubit state alone can be expressed as[24]

$$\dot{\rho}_{11} = -2\sqrt{\Gamma_{ci}}\rho_{11}\rho_{22}\xi, \quad (5)$$

$$\begin{aligned}\dot{\rho}_{12} &= -(i\tilde{\Omega}_q + \Gamma_d)\rho_{12} + \sqrt{\Gamma_{ci}}\langle\sigma_z\rangle\rho_{12}\xi \\ &\quad + i\sqrt{\Gamma_{ba}}\rho_{12}\xi,\end{aligned} \quad (6)$$

for the diagonal and off-diagonal elements of the density matrix, respectively. Here we have also absorbed a generalized dynamic ac-Stark shift, $B(t) = 2\chi \, \text{Re}[\alpha_1(t)\alpha_2^*(t)]$ into the effective frequency of the qubit, $\widetilde{\Omega}_q$.

Eqs.(5) and (6) are defined on the basis of Itó calculus and is the form for numerical simulations[25,26]. However, as to be clear in the following, in order to get the correct result of the quantum Bayesian rule, one needs to convert it into the Stratonovich-type form. We thus have

$$\dot{\rho}_{11} = -2\sqrt{\Gamma_{ci}}\rho_{11}\rho_{22}I(t), \tag{7}$$

$$\begin{aligned}\dot{\rho}_{12} &= -(i\widetilde{\Omega}_q + \Gamma_d)\rho_{12} + \frac{1}{2}(\Gamma_{ba} + \Gamma_{ci})\rho_{12} \\ &\quad + \sqrt{\Gamma_{ci}}\langle\sigma_z\rangle_t \rho_{12}I(t) + i\sqrt{\Gamma_{ba}}\rho_{12}I(t).\end{aligned} \tag{8}$$

We see that, compared to the Itó type Eqs. (5) and (6), the Gaussian noise $\xi(t)$ in all the noisy terms has been replaced now by the homodyne current $I(t)$ which is given by Eq. (1).

Now we are ready to integrate Eqs. (7) and (8). For the first equation of $\rho_{11}(t)$, perform $\int_0^{t_m} dt\,(\cdots)$ on both sides. Noting that $\int_0^{t_m} d\rho_{11}/(\rho_{11}\rho_{22}) = \ln(\rho_{11}/\rho_{22})|_0^{t_m}$, we obtain $\rho_{11}(t_m)/\rho_{22}(t_m) = [\rho_{11}(0)/\rho_{22}(0)]e^{-2X(t_m)}$, where $X(t_m) = \int_0^{t_m} dt \sqrt{\Gamma_{ci}(t)}I(t)$. In deriving, we have used the property $\rho_{11} + \rho_{22} = 1$. Using this property again, we may split the solution into two equations

$$\rho_{11}(t_m) = \rho_{11}(0)e^{-X(t_m)}/N(t_m), \tag{9}$$

$$\rho_{22}(t_m) = \rho_{22}(0)e^{X(t_m)}/N(t_m), \tag{10}$$

where $N(t_m) = \rho_{11}(0)e^{-X(t_m)} + \rho_{22}(0)e^{X(t_m)}$. Further, integrating the second equation for $\rho_{12}(t)$, we obtain

$$\begin{aligned}\rho_{12}(t_m) &= \rho_{12}(0)e^{-i\int_0^{t_m} dt \widetilde{\Omega}_q}e^{-\int_0^{t_m} dt(\Gamma_d - \Gamma_m/2)} \\ &\quad \times e^{-i\int_0^{t_m} dt(-\sqrt{\Gamma_{ba}})I(t)}e^{\int_0^{t_m} dt\sqrt{\Gamma_{ci}}I(t)\langle\sigma_z\rangle_t}.\end{aligned} \tag{11}$$

In this solution, we introduced the measurement rate $\Gamma_m = \Gamma_{ba} + \Gamma_{ci}$; and all the factors except the last one have been in explicit forms of integration with known integrands. For the last factor, since $\int_0^{t_m} dt\sqrt{\Gamma_{ci}(t)}I(t)\langle\sigma_z\rangle_t = \int_0^{t_m} dX(t)[2\rho_{11}(t) - 1]$, substituting the explicit solution of $\rho_{11}(t)$ to complete the integration yields

$$\exp[\int_0^{t_m} dt\sqrt{\Gamma_{ci}(t)}\,I(t)\,\langle\sigma_z\rangle_t] = [\rho_{11}(0)e^{-X(t_m)} + \rho_{22}(0)e^{X(t_m)}]^{-1}. \tag{12}$$

Then we reexpress the solution of the off-diagonal element $\rho_{12}(t_m)$ in a compact form as

$$\rho_{12}(t_m) = [\rho_{12}(0)/N(t_m)]D(t_m)e^{-i[\Phi_1(t_m)+\Phi_2(t_m)]}, \tag{13}$$

where we have introduced

$$D(t_m) = \exp\left\{-\int_0^{t_m} dt\,[\Gamma_d(t) - \Gamma_m(t)/2]\right\} \tag{14}$$

$$\Phi_1(t_m) = \int_0^{t_m} dt\,\widetilde{\Omega}_q(t), \tag{15}$$

$$\Phi_2(t_m) = -\int_0^{t_m} dt\sqrt{\Gamma_{ba}(t)}\,I(t). \tag{16}$$

Desirably, these factors recover those proposed in ref. 27 from different insight and analysis. As demonstrated in ref. 27, these factors have important effects to correct the "bare" Bayesian rule for the off-diagonal elements. As to be seen soon (after simple algebra), the above treatment provides also a *reliable* method allowing us to obtain new and precise expressions for the prior distribution knowledge of the output currents which are of essential importance to the Bayesian inference for the cQED measurement.

To make the results derived above in the standard form of Bayesian rule, let us rewrite $e^{-X(t_m)}/N(t_m) = P_1(t_m)/N(t_m)$ and $1/N(t_m) = \sqrt{P_1(t_m)P_2(t_m)}/N(t_m)$, where $N(t_m) = \rho_{11}(0)P_1(t_m) + \rho_{22}(0)P_2(t_m)$ and the current distribution probabilities read

$$P_{1(2)}(t_m) = \exp\left\{-\langle[I(t) - \bar{I}_{1(2)}(t)]^2\rangle_{t_m}/(2V)\right\}, \tag{17}$$

where $\bar{I}_{1(2)}(t) = \mp\sqrt{\Gamma_{ci}(t)}$ and $\langle\bullet\rangle_{t_m} = (t_m)^{-1}\int_0^{t_m} dt\,(\bullet)$, and $V = 1/t_m$ characterizes the distribution variance. For Bayesian inference, the "prior" knowledge of the distribution probabilities $P_{1(2)}(t_m)$ associated with qubit state $|1\rangle (|2\rangle)$), should be known in advance. Then, one utilizes them to update the measured state based on the collected output currents, using the following *exact* quantum Bayesian rule. First, for the off-diagonal element,

$$\begin{aligned} \rho_{12}(t_m) &= \rho_{12}(0)[\sqrt{P_1(t_m)P_2(t_m)}/N(t_m)] \\ &\quad \times D(t_m)\exp\{-i[\Phi_1(t_m) + \Phi_2(t_m)]\}. \end{aligned} \qquad (18)$$

Second, for the diagonal elements,

$$\rho_{jj}(t_m) = \rho_{jj}(0)P_j(t_m)/N(t_m), \qquad (19)$$

where $j = 1, 2$. The results of Eq. (14)–(19) constitute the main contribution of this work.

**Result of simpler case.** The original work of quantum Bayesian approach considered a charge qubit ($H_q = \frac{\omega_q}{2}\sigma_z$) measured by point-contact detector[28]. This problem can be described by the following QTE which has been applied broadly in quantum optics[1,2]

$$\dot{\rho} = -i[H_q, \rho] + \gamma' D[\sigma_z]\rho + \sqrt{\gamma}H[\sigma_z]\rho\xi(t). \qquad (20)$$

In this equation the Lindblad term takes $D[\sigma_z]\rho = \sigma_z\rho\sigma_z - \rho$, and the information-gain backaction term reads $H[\sigma_z]\rho = \sigma_z\rho + \rho\sigma_z - 2\text{Tr}[\sigma_z\rho]\rho$. $\gamma'$ and $\gamma$ are respectively the measurement decoherence and information-gain rates (constants). And, this equation is conditioned on the measurement current $I(t) = 2\sqrt{\gamma}\,\text{Tr}[\sigma_z\rho(t)] + \xi(t)$, since both share the same stochastic noise $\xi(t)$.

Following precisely the same procedures of solving the above cQED system, one can arrive to similar results as Eqs. (9)(10) and (13) with, however, simpler $X(t_m)$ which is now given by $X(t_m) = -2\sqrt{\gamma}\int_0^{t_m} dt I(t)$, owing to the constant $\gamma$. Also, now the dephasing factor simply reads $D(t_m) = e^{-\tilde{\gamma}t_m}$ where $\tilde{\gamma} = \gamma' - \gamma$. For this simpler setup, the phase factor $\Phi_1$ takes the trivial result $\Phi_1(t_m) = \omega_q t_m$ and the factor $\Phi_2$ vanishes. The simpler result of $X(t_m)$ allows us to introduce, from the factors $e^{\pm X(t_m)}$ in Eqs. (9)(10) and (13), the distribution probabilities of the standard Gaussian form

$$P_{1(2)}(t_m) = (2\pi V)^{-1/2}\exp\left[-\left(I_m - \bar{I}_{1(2)}\right)^2/(2V)\right], \qquad (21)$$

where $I_m = (t_m)^{-1}\int_0^{t_m} dt I(t)$ and $\bar{I}_{1(2)} = \pm 2\sqrt{\gamma}$. With these identifications, one recovers the quantum Bayesian rule proposed by Korotkov in ref. 28.

Similarly, for the cQED setup, from $I(t) = -\sqrt{\Gamma_{ci}(t)}\,\text{Tr}[\sigma_z\rho(t)] + \xi(t)$, simple experience likely tells us that, corresponding to the qubit states $|1\rangle$ and $|2\rangle$, the stochastic current $I_m$ (average of the stochastic $I(t)$ over $t_m$) should be respectively centered at $\bar{I}_{1(2)} = \mp(t_m)^{-1}\int_0^{t_m} dt\sqrt{\Gamma_{ci}(t)}$, in terms of the Gaussian distribution as given by Eq. (21). However, out of our expectation, this is not true. Below we display numerical results to show the *exactness* of the Bayesian rule Eqs. (18) and (19) when associated with (16), rather than with the Gaussian formula (20).

**Numerical results.** To implement the above *exact* Bayesian rule for state inference in practice, we should carry out in advance the rates $\Gamma_{ci}(t), \Gamma_{ba}(t), \Gamma_d(t)$ and as well the ac-Stark shift $B(t)$. From the expressions of these quantities, we know that the key knowledge required is the coherent-state parameters $\alpha_1(t)$ and $\alpha_2(t)$ which are, respectively, the consequence of the interplay of external driving and cavity damping for qubit states $|1\rangle$ and $|2\rangle$, satisfying $\dot{\alpha}_{1,2}(t) = -i\varepsilon_m - i\widetilde{\Delta}_r^{(\mp)}\alpha_{1,2}(t)$, where $\widetilde{\Delta}_r^{(\mp)} = (\Delta_r \pm \chi) - i\kappa/2$. The simple analytic solutions read

$$\alpha_{1,2}(t) = \bar{\alpha}_{1,2}[1 - e^{-i\widetilde{\Delta}_r^{(\mp)}t}] + \alpha_0 e^{-i\widetilde{\Delta}_r^{(\mp)}t}. \qquad (22)$$

In this solution, $\bar{\alpha}_{1,2} = -\varepsilon_m/\widetilde{\Delta}_r^{(\mp)}$ are the steady-state cavity fields; and $\alpha_0$ is the initial cavity field before measurement which is zero if we start with a vacuum.

In Fig. 2, taking a specific trajectory for each case as an example, we compare the results from different approaches. In this plot we adopt a reduced system of units by scaling energy and frequency with the microwave drive strength ($\varepsilon_m$). We have assumed two types of parameters in the simulation: for the upper row (weaker response) we assumed $\chi = 0.1$ while for the lower one (stronger response) we assumed $\chi = 0.5$, both violating the joint condition of "bad"-cavity and weak-response since we commonly used $\kappa = 2$. (Other parameters are referred to the figure caption). The basic requirement of the "bad"-cavity and weak response limits is $\kappa \gg \chi$. This has a couple of consequences: (i) The time-dependent factor in Eq. (22), $e^{\pm i\chi t}e^{-\kappa t/2}$, would become less important so that we can neglect the transient dynamics of the cavity field and all the rates ($\Gamma_d, \Gamma_{ba}$ and $\Gamma_{ci}$) can be treated as (steady state) constants[16]. (ii) It makes the measurement rate ($\Gamma_m$) much smaller than $\kappa$, and the purity factor $D(t_m)$ almost unity. This means that, with respect to the cavity photon's leakage, the (gradual collapse) measurement process is slow, and the qubit state remains almost pure in the ideal case (in the absence of photon loss and amplification noise). One can check that the parameters used in Fig. 2 (especially the case $\chi = 0.5$) violate these criteria.

We find that for both cases the results from our exact Bayesian rule, Eqs. (18) and (19) together with (16), precisely coincides with those from simulating the QTE. As a comparison, in Fig. 2 we plot also the results from other two approximate Bayesian approaches. One is the Bayesian approach constructed by Korotkov in ref. 16 under the bad-cavity and weak-response limits, which is labeled in Fig. 2 by "K". Another is the Bayesian rule obtained in ref. 27, which holds the same factors as shown in Eqs. (14)–(16) but involves the usual Gaussian distribution of the type of Eq. (21). The results are labeled by "G" in Fig. 2. We find from this plot that, for the modest violation of the bad-cavity and weak-response limits, the "G" results are better than the "K" ones. However, for strong violation, the "G" results will become unreliable as well. In contrast, as we observe here and have checked for many more trajectories in the case of violating the bad-cavity and weak-response limits, the exact Bayesian rule can always work precisely while both the "K" and "G" approaches failed.

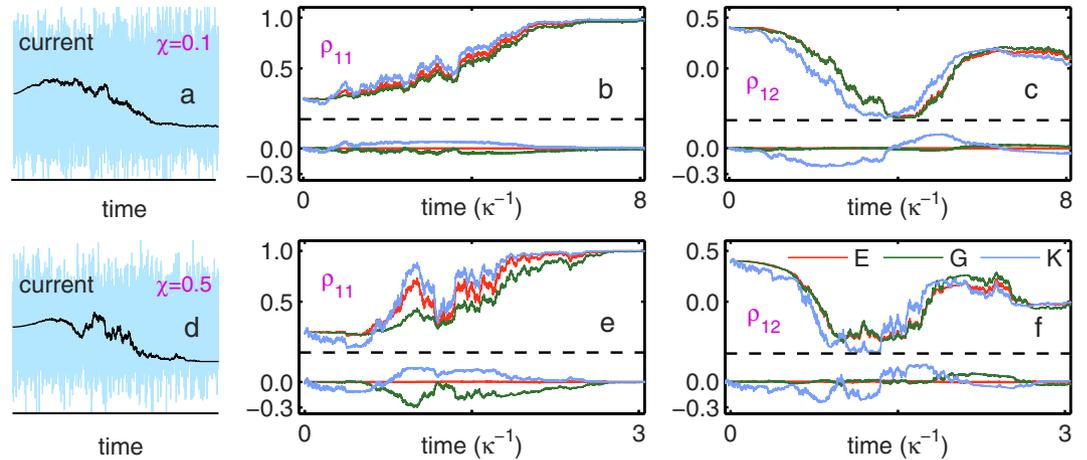

**Figure 2. Accuracy demonstration of the Bayesian rule against the quantum trajectory equation. (a,d)** Stochastic currents in the homodyne detection for dispersive coupling $\chi=0.1$ and 0.5. The black curves denote the coarse-grained results for visual purpose, while the original currents (blue ones) are actually used for state estimate (inference). **(b,c)** State (density matrix) evolution under continuous measurement for a relatively weak qubit-cavity coupling, $\chi=0.1$. **(e,f)** State evolution under continuous measurement for a strong qubit-cavity coupling, $\chi=0.5$. **(b,c,e,f)** The curves "E" (red), "G" (green), and "K" (blue) denote, respectively, our exact Bayesian rule, Eqs. (18) and (19) together with (17), the approximate one involving instead the usual Gaussian distribution of Eq. (21), and that proposed by Korotkov[16] under the bad-cavity and weak-response limits. In each figure, the lower panel plots also the *difference* from the quantum trajectory equation result, indicating that the BR proposed in this work is indeed exact. In all these numerical simulations, we chose the LO's phase $\varphi=\pi/4$ and adopted a system of reduced units with parameters $\Delta_r=0$, $\varepsilon_m=1.0$, and $\kappa=2.0$.

## Discussion

To summarize, for continuous weak measurements in circuit-QED, we carry out the analytic and *exact* solution of the effective QTE which generalizes the quantum Bayesian approach developed in ref. 16 and applied in circuit-QED experiments[13–15]. The new result is not bounded by the "bad"-cavity and weak-response limits as in ref. 16, and improves the quantum Bayesian rule via amending the distribution probabilities of the output currents and several important phase factors[16,27].

The efficiency of quantum Bayesian approach can be understood with the illustrative Fig. 1. Instead of the successive step-by-step state estimations (over many infinitesimal time intervals of *dt*) using QTE, the Bayesian rule offers the great advantage of *one-step* estimation, with a major job by inserting the continuous measurement outcomes (currents) into the simple expressions of the prior knowledge $P_{1,2}(t_m)$ given by Eq. (17) and the phase correction factor $\Phi_2(t_m)$. Other $\alpha_{1,2}(t)$-dependent factors are independent of the measurement outcomes and can be carried out in advance. The time integration in $P_{1,2}(t_m)$ and $\Phi_2(t_m)$ can be completed as soon as the continuous measurement over $(0, t_m)$ is finished. Therefore, the state estimation can be accomplished using Eqs. (18) and (19), with similar efficiency as using the usual simple Gaussian distribution Eq. (21).

In this work we have focused on the single quadrature measurement. For the so-called $(I, Q)$ two quadrature measurement, following the same procedures of this work or even simply based on the characteristic structure of the final results, Eqs. (14)–(19), one can obtain the associated quantum Bayesian rule as well[16,27].

We finally remark that, just as Bayesian formalism is the central theory for noisy measurements in classical control and information processing problems, quantum Bayesian approach can be taken as an efficient theoretical framework for continuous quantum weak measurements. Compared to the "construction" method[16], the present work provides a direct and more reliable route to formulate the quantum Bayesian rule. Applying similar method to more complicated cases such as joint measurement of multiple qubits or to other systems is of interest for further studies.

## Methods

**Circuit-QED setup and measurements.** Under reasonable approximations, the circuit-QED system resembles the conventional atomic cavity-QED system which has been extensively studied in quantum optics. Both can be well described by the Jaynes-Cummings Hamiltonian. In dispersive regime[3–5], i.e., the detuning between the cavity frequency ($\omega_r$) and qubit energy ($\omega_q$), $\Delta=\omega_r-\omega_q$, being much larger than the coupling strength $g$, the system Hamiltonian (in the rotating frame with the microwave driving frequency $\omega_m$) reads[3–5]

$$H_{\text{eff}} = \Delta_r a^\dagger a + \frac{\widetilde{\omega}_q}{2}\sigma_z + \chi a^\dagger a \sigma_z + (\varepsilon_m^* a + \varepsilon_m a^\dagger),\tag{23}$$

where $\Delta_r = \omega_r - \omega_m$ and $\widetilde{\omega}_q = \omega_q + \chi$, with $\chi = g^2/\Delta$ the dispersive shift of the qubit energy. In Eq. (23), $a^\dagger$ ($a$) and $\sigma_z$ are respectively the creation (annihilation) operator of cavity photon and the quasi-spin operator (Pauli matrix) for the qubit. $\varepsilon_m$ is the microwave drive amplitude for continuous measurements. For single-quadrature (with reference local oscillator phase $\varphi$) homodyne detection of the cavity photons, the measurement output can be expressed as $I(t) = \sqrt{\kappa}\,\mathrm{Tr}[(ae^{-i\varphi} + a^\dagger e^{i\varphi})\rho(t)] + \xi(t)$. After eliminating the cavity degrees of freedom via the "polaron"-transformation, this output current turns to Eq. (1). Moreover, the conditional qubit-cavity joint state $\rho(t)$, which originally satisfies the standard optical quantum trajectory equation[1,2], becomes now after eliminating the cavity states[24]:

$$\begin{aligned}\dot\rho = &-i\frac{\widetilde{\omega}_q + B(t)}{2}[\sigma^z,\rho] + \frac{\Gamma_d(t)}{2}\mathcal{D}[\sigma^z]\rho\\ &-\sqrt{\Gamma_{ci}(t)}\,\mathcal{M}[\sigma^z]\rho\,\xi(t) + i\frac{\sqrt{\Gamma_{ba}(t)}}{2}[\sigma^z,\rho]\xi(t).\end{aligned}\tag{24}$$

In addition to the Lindblad term (the second one on the r.h.s), the other superoperator is introduced through $\mathcal{M}[\sigma^z]\rho = (\sigma^z\rho + \rho\sigma^z)/2 - \langle\sigma^z\rangle\rho$, where $\langle\sigma^z\rangle = \mathrm{Tr}[\sigma^z\rho]$. In this result, $B(t)$ is the dynamic ac-Stark shift and $\Gamma_d$, $\Gamma_{ci}$ and $\Gamma_{ba}$ are, respectively, the overall measurement decoherence, information gain, and no-information back-action rates (with explicit expressions given by Eqs. (2)–(4)). In the qubit-state basis, Eq. (24) gives Eqs. (5) and (6).

**Conversion rule.** In order to convert Eqs. (5) and (6) to Eqs. (7) and (8), in this Appendix we specify the conversion rule from Itó to Stratonovich stochastic equations[25,26]. Suppose for instance we have a set of Itó-type stochastic equations

$$\left[\dot Y_j\right]_I = G_j(t) + F_j(t)\xi(t),\tag{25}$$

with $j = 1, 2, \cdots, K$. The corresponding Stratonovich-type equations read

$$\left[\dot Y_j\right]_S = \left[\dot Y_j\right]_I - \frac{1}{2}\sum_{i=1}^K F_i\frac{\partial F_j}{\partial Y_i}.\tag{26}$$

Comparing Eq. (25) with Eqs. (5) and (6), we identify $Y_1 = \rho_{11}$ and $Y_2 = \rho_{12}$. Then we have $F_1 = -2\sqrt{\Gamma_{ci}}Y_1(1 - Y_1)$, and $F_2 = [\sqrt{\Gamma_{ci}}(2Y_1 - 1) + i\sqrt{\Gamma_{ba}}]Y_2$. Applying the conversion rule Eq. (26), one can convert Eqs. (5) and (6) to Eqs. (7) and (8) by completing the following algebraic manipulations:

$$\begin{aligned}\dot\rho_{11} &= -2\sqrt{\Gamma_{ci}}\rho_{11}\rho_{22}\xi + \sqrt{\Gamma_{ci}}\rho_{11}\rho_{22}[-2\sqrt{\Gamma_{ci}}(1 - 2\rho_{11})]\\ &= -2\sqrt{\Gamma_{ci}}\rho_{11}\rho_{22}(-\sqrt{\Gamma_{ci}}\langle\sigma_z\rangle + \xi)\\ &= -2\sqrt{\Gamma_{ci}}\rho_{11}\rho_{22}I(t),\end{aligned}\tag{27}$$

and

$$\begin{aligned}\dot\rho_{12} &= -\left(i\widetilde{\Omega}_q + \Gamma_d\right)\rho_{12} + (\sqrt{\Gamma_{ci}}\langle\sigma_z\rangle + i\sqrt{\Gamma_{ba}})\rho_{12}\xi\\ &\quad -\frac{1}{2}(\sqrt{\Gamma_{ci}}\langle\sigma_z\rangle + i\sqrt{\Gamma_{ba}})(\sqrt{\Gamma_{ci}}\langle\sigma_z\rangle + i\sqrt{\Gamma_{ba}})\rho_{12}\\ &\quad -\frac{1}{2}(2\sqrt{\Gamma_{ci}}\rho_{12})(-2\sqrt{\Gamma_{ci}}\rho_{11}\rho_{22})\\ &= -\left(i\widetilde{\Omega}_q + \Gamma_d - \Gamma_m/2\right)\rho_{12}\\ &\quad + (\sqrt{\Gamma_{ci}}\langle\sigma_z\rangle + i\sqrt{\Gamma_{ba}})\rho_{12}I(t).\end{aligned}\tag{28}$$

In the above manipulations, we have used $\langle\sigma_z\rangle = \rho_{11} - \rho_{22} = 2\rho_{11} - 1$, $1 - \langle\sigma_z\rangle^2 = 4\rho_{11}\rho_{22}$, and $\Gamma_m = \Gamma_{ci} + \Gamma_{ba}$.

### Acknowledgements

This work was supported by the NNSF of China under grants No. 101202101 & 10874176, and the State "973" Project under grants No. 2011CB808502 & 2012CB932704.

### Author Contributions

X.Q.L. supervised the work; W.F., P.F.L. and L.P.Q. carried out the calculations; X.Q.L. wrote the paper and all authors reviewed it. W.F. and P.F.L. equally contributed to this work.

### Additional Information

**Competing financial interests:** The authors declare no competing financial interests.

**How to cite this article**: Feng, W. *et al.* Exact quantum Bayesian rule for qubit measurements in circuit QED. *Sci. Rep.* **6,** 20492; doi: 10.1038/srep20492 (2016).